\documentclass{aa}
\usepackage{graphicx}
\begin{document}

%%%%%%%%%%%%%%%%%%%%%%%%%%%%%%%%%%%%%%%%
\title{Spiral Structure when Setting up Pericentre Glow:
  Possible Giant Planets at Hundreds of AU in the HD141569 Disk}
%\subtitle{}
\author{M. C. Wyatt}%\inst{1}}
\offprints{M. C. Wyatt}
\institute{UK Astronomy Technology Centre, Royal Observatory,
             Edinburgh EH9 3HJ, UK\\
             \email{wyatt@roe.ac.uk} }
\date{Submitted 9 May 2005; accepted 3 June 2005}
\authorrunning{Wyatt}
\titlerunning{Spiral Structure when Setting up Pericentre Glow}

%%%%%%%%%%%%%%%%%%%%%%%%%%%%%%%%%%%%%%%%
\abstract{
This paper discusses the impact of introducing a planet on an eccentric
orbit into a dynamically cold planetesimal disk.
That planet's secular perturbations cause the orbits of the planetesimals
to evolve in such a way that at any one time planetesimals at the same
distance from the star have common pericentres and eccentricities.
This causes the surface density distribution of an extended planetesimal
disk to exhibit two spirals, one exterior the other interior to
the planet's orbit.
These two spirals unwind in different directions and their structure is
described by just two parameters:
the time since the planet was introduced relative to the characteristic secular
timescale, $t_{sec(3:2)} = 0.651\sqrt{a_{pl}^3/M_\star}(M_\star/M_{pl})$;
and the planet's eccentricity, $e_{pl}$.
At late times the spirals become tightly wound and the offset centre of
symmetry of the pericentre glow approximation is recovered.
Comparison with spiral structure seen in the HD141569 disk shows that
its spiral at 325 AU is similar to the structure that would be caused by
introducing a planet into the disk 5 Myr ago with a mass in the range
$0.2-2M_{Jup}$ orbiting at 235-250 AU with an eccentricity of 0.05-0.2;
likewise a Saturn mass planet at 150 AU would cause structure like that
seen at 200 AU.
More definitive statements about any planets orbiting HD141569 from
this model could be made once the effect of the binary companion on the
disk is known (e.g., from knowledge of its orbit), and once the disk's
structure has been better characterised down to 100 AU, including the
location of the star within the disk.
The relatively young age of this system ($\sim 5$ Myr) means that
if giant planets really do exist at hundreds of AU from HD141569,
this provides a unique opportunity to set constraints on the
mechanism by which those planets came to be at such large distances,
especially since the structure of the disk out of which those planets
would have formed can be imaged.
}
%%%%%%%%%%%%%%%%%%%%%
\maketitle

%%%%%%%%%%%%%%%%%%%%%
\section{Introduction}
\label{s:intro}
Circumstellar disks play a vital role in our understanding of how planetary systems
form and evolve.
It is within such disks that planets are thought to form on timescales
of $\sim 10$ Myr.
Imaging of the disks thus shows us where the material is that could either end up
forming planets, or for older systems where the debris is after planet formation
has already taken place (Backman \& Paresce 1993).
Disk images also often show considerable structure ranging from clumps,
brightness asymmetries, warps and spirals (Holland et al. 1998; Greaves et al. 1998;
Telesco et al. 2000; Heap et al. 2000; Clampin et al. 2003).
Ascertaining the origin of these structures is particularly important because
they may be indicative of the status of planet formation in these systems.
It is even possible to use these structures to infer information about planets which
have already formed, but which are not directly detectable themselves, since
the structures may be the consequence of dynamical perturbations to
the disk from a relatively small planet (Wyatt et al. 1999; Ozernoy et al. 2000;
Wyatt 2003; Deller \& Maddison 2005).
It is thus necessary to understand the various ways in which planetary perturbations
can cause structure in disks.

It is especially important to consider structures in intermediate age systems
($\sim 10$ Myr), since this is the time when planets should already be largely
formed, but before the planetary system has settled to its final configuration.
This is also the time when the structures seen in older disks
are thought to form.
Disks at this age have the benefit of being relatively bright compared
with their older counterparts (Spangler et al. 2001), but not so dense as to be
optically thick like their younger counterparts which complicates interpretation
of disk structure.
HD141569 is a good example of a disk around such an intermediate aged star:
it has an age of $\sim 5$ Myr (Weinberger et al. 2000; Mer\'{i}n et al. 2004)
and a fractional infrared luminosity of $f = L_{ir}/L_\star = 0.018$
(Mer\'{i}n et al. 2004).
Starlight scattered by $\mu$m-sized dust in its disk has also been resolved
in optical and near-IR coronagraphic images revealing considerable structure:
the disk density peaks in rings at 200 and 325 AU (Augereau et al. 1999;
Weinberger et al. 1999; Mouillet et al. 2001) which are tightly wound spiral
structures (Clampin et al. 2003);
and diffuse dust is seen up to 1200 AU in a more open spiral arm
structure (Mouillet et al. 2001; Clampin et al. 2003).
Mid-IR emission from this disk has also been marginally resolved at
$<200$ AU (Fisher et al. 2000; Marsh et al. 2002) which is shown to be
a region of lower density than the outer disk.
Gas has also been detected in this inner region (Zuckerman, Forveille
\& Kastner 1995; Brittain \& Rettig 2002).

In three recent papers, it was shown that some of the structure in the
outer disk could be caused by the gravitational perturbations of a binary
companion (Augereau \& Papaloizou 2004; Quillen et al. 2005; Ardila et
al. 2005).
The orbit of the two coeval M stars seen at $\sim 1000$ AU projected
separation is at present unknown (Weinberger et al. 1999).
However, Augereau \& Papaloizou showed how secular perturbations from these
stars could explain the observed spiral structure and azimuthal
asymmetry of the ring at 325 AU if the orbit had a suitable eccentricity
and pericentre orientation.
Quillen et al. (2005) used hydrodynamic simulations to show
how tidal perturbations from the same binary companion could explain the
more extended outer spiral arms at $>400$ AU.
Their model indicated a significantly different orbit for the companion
to that of Augereau \& Papaloizou, notably in the orientation of its
periastron, and could not account for the structure of the ring at
325 AU.
Ardila et al. (2005) considered the possibility that
the disk structure is caused by a flyby encounter between the disk
and the binary companion, with similar conclusions to Quillen et al.,
that such an event could explain the open spiral arms at $>400$ AU,
but not reproduce the tightly wound structures at 200 and 325 AU.

This paper describes an alternative cause for the tightly wound spiral
structures at 200 and 325 AU which is very similar in concept to the
model of Augereau \& Papaloizou (2004):
instead of invoking the long term (secular) gravitational perturbations of
a binary companion on a highly eccentric orbit, this paper invokes the 
secular perturbations of a moderately eccentric, relatively low mass planet
orbiting within the disk.
It was shown in Wyatt et al. (1999) how on long timescales, such perturbations
impose a forced eccentricity on the disk,
and how this causes one side of the disk to be hotter and brighter than the
other side, an effect they called pericentre glow.
They showed that this may be the cause for the lobe brightness asymmetry
observed in the HR4796 disk (Telesco et al. 2000).
This paper discusses the early stages of pericentre glow, i.e., on timescales
over which it is imposed on a disk.

Section \ref{s:sp} discusses the dynamics of secular perturbations and
shows the impact of introducing a planet on an eccentric orbit 
into an initially axisymmetric disk for the dynamical structure of that disk.
This section also demonstrates how the analytically derived orbital element
distribution causes spiral structure in the spatial distribution of
the planetesimals.
In section \ref{s:hd} the model is applied to the structure of HD141569.
The conclusions are given in section \ref{s:conc}.

%%%%%%%%%%%%%%%%%%%%%
\section{Secular Perturbations: Early Stages of Pericentre Glow}
\label{s:sp}
The gravitational forces from a planetary system that act to perturb the
orbit of a particle in the system can be decomposed into the sum of many
terms that are described by the particle's disturbing function, $R$.
The long-term average of these forces are the system's secular
perturbations, and the terms of the disturbing function that contribute
to these secular perturbations, $R_{sec}$, can be identified as those that
do not depend on the mean longitudes of either the planets or the particle
(the other forces having periodic variations).

%%%%%%%%%%%%%%%%%%%%%
\subsection{Restricted Three Body Problem}
\label{ss:rtbp}
Consider a particle that is orbiting a star of mass $M_\star$, that also
has a planet of mass $M_{pl}$ orbiting it.
The particle's orbit is described by the elements $a$, $e$, $\tilde{\omega}$.
To second order in eccentricities, the secular terms in the
particle's disturbing function are given by
(Brouwer \& Clemence 1961; Murray \& Dermott 1999; Wyatt et al.
1999):
\begin{equation}
  R_{sec} =  na^2 \left[ 0.5Ae^2
    + A_{pl} e e_{pl} cos(\tilde{\omega}-\tilde{\omega}_{pl}) \right],
\label{eq:rsec}
\end{equation}
where $n = (2\pi/t_{year})\sqrt{(M_\star/M_\odot)(a_\oplus/a)^3}$
is the mean motion of the particle in rad/s,
$t_{year} = 2\pi / \sqrt{GM_\odot/a_\oplus^3} = 3.156\times 10^7$ s is one year
measured in seconds, $a_\oplus = 1$ AU, and
  \begin{eqnarray}
    A   & = & + 0.25n \left( \frac{M_{pl}}{M_\star}     \right)
                \alpha_{pl}\overline{\alpha}_{pl} b_{3/2}^{1}(\alpha_{pl}),
    \label{eq:a} \\
    A_{pl} & = & -0.25n \left( \frac{M_{pl}}{M_\star}     \right)
                \alpha_{pl}\overline{\alpha}_{pl} b_{3/2}^{2}(\alpha_{pl}),
    \label{eq:aj}
  \end{eqnarray}
where $\alpha_{pl} = a_{pl}/a$ and $\overline{\alpha}_{pl} = 1$ for $a_{pl}<a$, and 
$\alpha_{pl} = \overline{\alpha}_{pl} = a/a_{pl}$ for $a_{pl}>a$,
and $b_{3/2}^s(\alpha_{pl}) = (\pi)^{-1}\int_0^{2\pi}(1-2\alpha_{pl}\cos{\psi}+
\alpha_{pl}^2)^{-3/2}\cos{s\psi}d\psi$ are the Laplace coefficients ($s=1,2$).
$A$ and $A_{pl}$ are in units of rad/s, and $R_{sec}$ is in units of m$^2$/s$^2$.

The effect of these perturbations on the orbital elements of the
particle can be found using Lagrange's planetary equations
(\cite{bc61}; \cite{md99}).
The semimajor axis of the particle remains constant, 
$\dot{a}_{sec} = 0$, while the variation of its eccentricity
is best described when coupled with the
variations of its longitude of pericentre
using the complex eccentricity variable, $z$:
\begin{equation}
  z = e*\exp{i\tilde{\omega}}, \label{eq:zdefn}
\end{equation}
where $i^2 = -1$.
Lagrange's planetary equations give the orbital
element variations due to secular perturbations as:
\begin{equation}
  \dot{z}_{sec} = iAz + A_{pl} z_{pl}, \label{eq:zdot}
\end{equation}
where $z_{pl}$ is the complex eccentricity of the planet, which
remains constant.

Equation (\ref{eq:zdot}) can be solved to give
the secular evolution of the particle's complex
eccentricity.
This evolution is decomposed into two distinct
elements --- the "forced", $z_f$, and "proper", $z_p$,
elements --- that are added vectorially in the complex plane:
\begin{eqnarray}
  z(t) = z_f + z_p(t) & = & \left[ b_{3/2}^{2}(\alpha_{pl}) /
    b_{3/2}^{1}(\alpha_{pl}) \right]
    e_{pl}\ast\exp{i\tilde{\omega}_{pl}} + \nonumber \\
   & & e_p\ast\exp{i(+At+\beta_0)},  \label{eq:z}
\end{eqnarray}
where $e_p$ and $\beta_0$ are determined by the particle's initial conditions.
A particle's forced eccentricity $z_f$ depends only on the
eccentricity of the planet's orbit, as well as on the ratio of its own semimajor
axis to that of the planet;
it is constant in time and independent of the planet's mass.
Thus the secular evolution of $z$ is to precess anticlockwise around
a circle centred on the forced eccentricity with a radius $e_p$ at a constant rate,
$A$.

\begin{figure}
  \centering
  \includegraphics[width=3in]{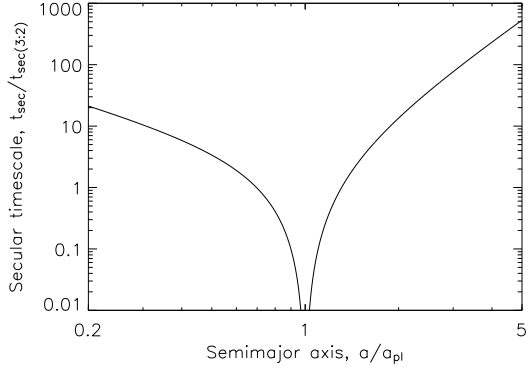}
  \caption{The secular precession timescale relative to that at a semimajor
  axis corresponding to the planet's outer 3:2 mean motion resonance (i.e.,
  $a = 1.31a_{pl}$).}
  \label{fig:tsecvsa}
\end{figure}

The secular precession timescale, $t_{sec} = 2\pi /At_{year}$, depends on both
the ratio of the masses of the planet and the star $\mu = M_{pl}/M_\star$, and
on the ratio of the semimajor axes of the particle and the planet ($\alpha$,
$\overline{\alpha}$).
Here I define a timescale, $t_{sec(3:2)}$ to be the precession timescale at a
semimajor axis corresponding to a planet's outer 3:2 mean motion resonance
(i.e., at $a/a_{pl} = 1.31$).
In this way, the precession timescale at different distances from the star
can be scaled using the following relations:
\begin{eqnarray}
  t_{sec}/t_{sec(3:2)} & = & 6.15 \alpha^{-2.5} \overline{\alpha}^2 / 
    b_{3/2}^1(\alpha_{pl}), \label{eq:tsec} \\
  t_{sec(3:2)}         & = & 0.651 t_{pl}/\mu, \label{eq:tsec32}
\end{eqnarray}
where $t_{pl} = \sqrt{(a_{pl}/a_\oplus)^3(M_\odot/M_\star)}$ is the
planet's orbital period in years.
The function $t_{sec}/t_{sec(3:2)}$ is plotted in Fig.~\ref{fig:tsecvsa}.

%%%%%%%%%%%%%%%%%%%%%
\subsection{Evolution of an Initially Cold Disk}
\label{ss:evcold}
\begin{figure}
  \centering
  \includegraphics[width=3in]{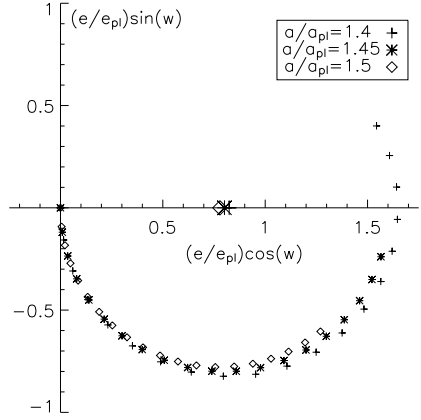}
  \caption{The secular evolution of the complex eccentricity,
  $z = e\ast \exp{i\tilde{w}}$, of planetesimals orbiting at
  semimajor axes of 1.4, 1.45 and $1.5a_{pl}$.
  Their eccentricities are shown relative to the eccentricity
  of the planet, $e_{pl}$.
  The large symbols on the x axis show the forced eccentricity about
  which these complex eccentricities are precessing.
  The precession is anticlockwise, and each point is advanced by
  a timestep of $0.05t_{sec(3:2)}$.
  The total evolution is $1t_{sec(3:2)}$.}
  \label{fig:zevolfew}
\end{figure}

Now consider the evolution of an initially cold planetesimal disk (i.e., on
orbits that are initially circular) after a planet on an eccentric orbit is
introduced into it.
The complex eccentricities of all planetesimals start at the origin ($e \approx 0$),
but then precess about the forced pericentre imposed on them.
Since the initial conditions are now defined, it is possible to see
that $e_p = e_f$ and $\beta_0 = \pi+\tilde{w}_{pl}$.
Planetesimals at different distances from the planet precess around circles of
different radii, and do so at different rates (e.g.,
Fig.~\ref{fig:tsecvsa}).
For example, Fig.~\ref{fig:zevolfew} shows the complex eccentricities of
planetesimals that are at $a=1.4, 1.45$ and $1.5a_{pl}$ at 20 equal timesteps
ending at a time of $1t_{sec(3:2)}$. 

\begin{figure*}
  \centering
  \begin{tabular}{c}
    \includegraphics[width=6.5in]{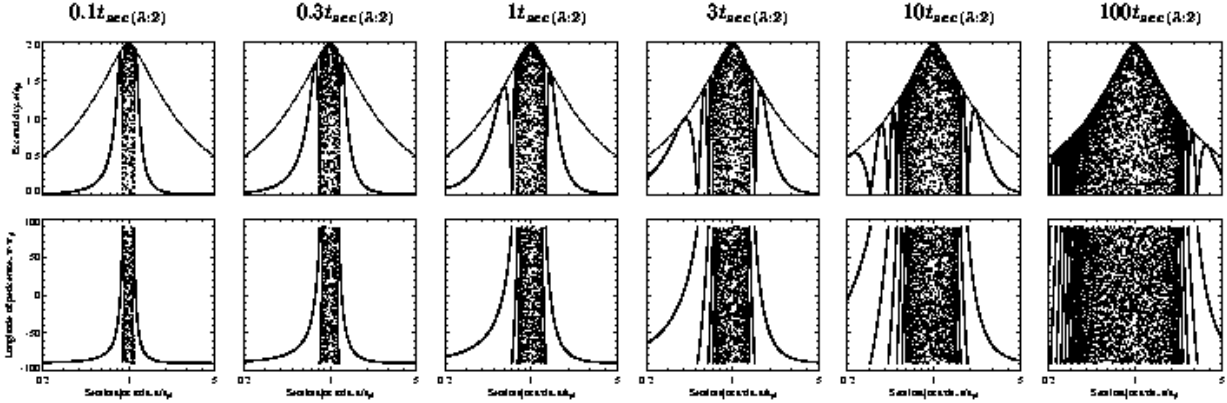}
  \end{tabular}
  \caption{The eccentricities (top) and pericentre orientations (bottom) of planetesimals
  in an initially dynamically cold planetesimal disk at times from left to right of
  0.1, 0.3, 1, 3, 10 and 100 $t_{sec(3:2)}$ after the introduction of a planet on an eccentric
  orbit.
  The eccentricities are shown relative to that of the planet, and
  the dotted lines on these plots indicate the maximum planetesimal eccentricity at a given
  distance from the star, $e_{max} = 2e_f$.
  The minima in the eccentricity plots occur at locations corresponding to an
  integer number of precession timescales;
  there is also a jump of pericentre orientation from $+90^\circ$ to $-90^\circ$
  at these locations.
  In the region close to the planet, several precession timescales
  have taken place leading to random eccentricities and pericentre orientations,
  although these still lie within the range $-90$ to $+90^\circ$.
  Movies showing the evolution of the planetesimals' eccentricities and pericentre
  orientations are available in the electronic edition of the journal.}
  \label{fig:ewvsa}
\end{figure*}

The instantaneous eccentricities and pericentre orientations of planetesimals
orbiting at different locations relative to the planet are shown in
Fig.~\ref{fig:ewvsa} for timescales of 0.1, 0.3, 1, 3, 10 and $100t_{sec(3:2)}$,
respectively.
Thus the dynamical structure of the disk can be determined uniquely from
the number of precession timescales $N_{sec(3:2)} = t/t_{sec(3:2)}$ which have been
completed since the planet was introduced.

%%%%%%%%%%%%%%%%%%%%%
\subsection{Spiral Structure}
\label{ss:ss}
\begin{figure*}
  \centering
  \begin{tabular}{c}
     \includegraphics[width=6.0in]{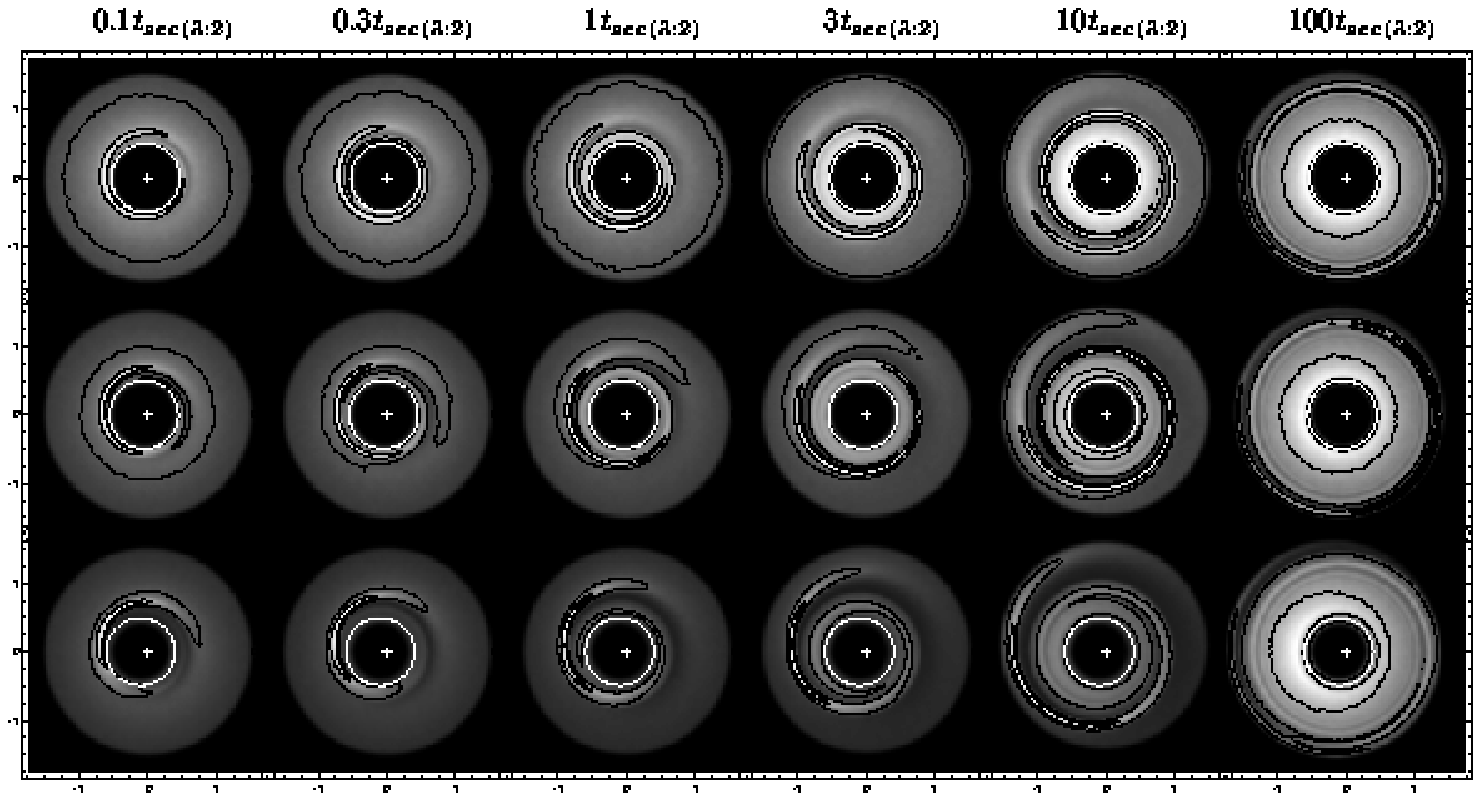}
  \end{tabular}
  \caption{The response of the spatial distribution of planetesimals orbiting with semimajor
  axes in the range $1-3a_{pl}$ to the introduction of a planet on an eccentric orbit.
  The different panels show the response for planet eccentricities of 0.05, 0.1, and 0.2
  (top to bottom), and at times of 0.1, 0.3, 1, 3, 10 and $100t_{sec(3:2)}$ (left to right).
  The x and y axes are scaled to the semimajor axis of the planet.
  The planet's orbit is shown by the white line with the pericentre direction to the right.
  The planetesimals are orbiting anticlockwise, but the direction of the planet's motion is
  unconstrained.
  The star is marked by the plus.
  The colour scale indicates the number of planetesimals in a given pixel (i.e., the surface density),
  and this number is also quantified by the black contours which indicate where the density is
  0.33 and 0.67 times the maximum value.
  A movie showing the evolution of the planetesimal distribution for $e_{pl} = 0.1$
  is available in the electronic edition of the journal.}
  \label{fig:pgspimn}
\end{figure*}

\begin{figure*}
  \centering
  \begin{tabular}{c}
     \includegraphics[width=6.0in]{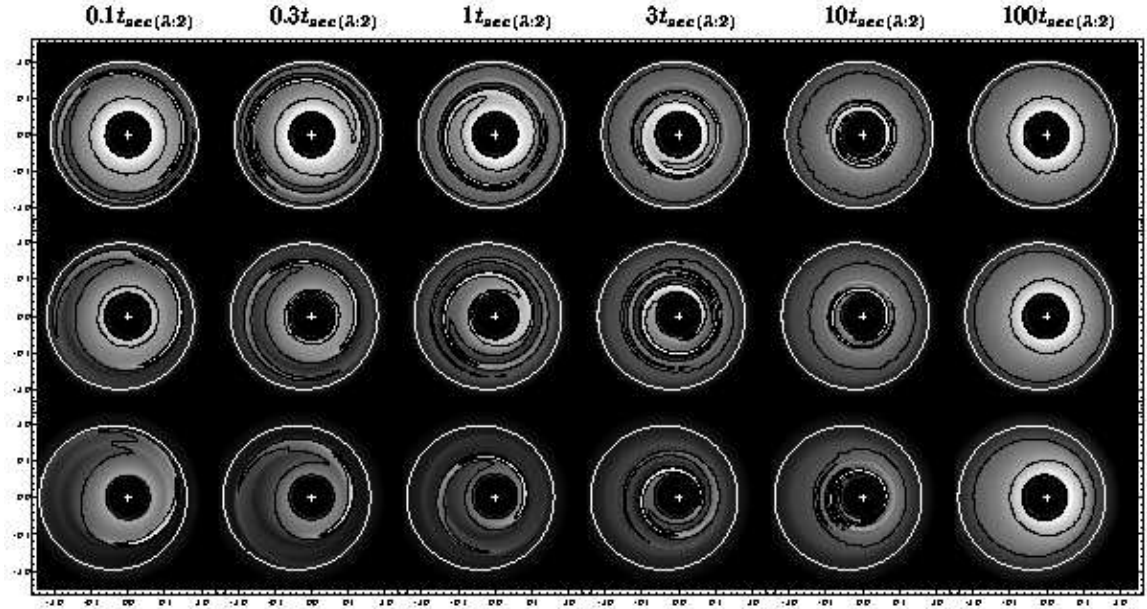}
  \end{tabular}
  \caption{The same as Fig.~\ref{fig:pgspimn} except for
  planetesimals orbiting with semimajor axes in the range $0.3-1a_{pl}$.
  Again, a movie showing the evolution of the planetesimal distribution for
  $e_{pl} = 0.1$ is available in the electronic edition of the journal.}
  \label{fig:pgspimnint}
\end{figure*}

The perturbation to the eccentricity distribution caused by the eccentric
planet affects the spatial distribution of the planetesimals:
it causes a spiral wave to form.
Figs.~\ref{fig:pgspimn} and \ref{fig:pgspimnint} show the spatial distribution of
planetesimals at different times 0.1, 0.3, 1.0, 3, 10 and $100t_{sec(3:2)}$
after the introduction of planets with eccentricities of 0.05, 0.1, and 0.2;
the two figures show the distribution of planetesimals exterior and interior to
the planet's orbit respectively and are derived from the analytically derived
dynamical structure of the disk shown in Fig.~\ref{fig:ewvsa} at the same times.
In these figures the planetesimals' semimajor axes were chosen randomly from an
appropriate range limited to 0.3-3 AU with a distribution given by $n(a) \propto
a^\delta$, where $\delta = 0$;
for each semimajor axis planetesimals were spread randomly (in mean
anomaly) around their orbits.

The response of the outer disk to the planet can be summarised as follows:
A narrow spiral forms which unwinds in the direction opposite to the planetesimals'
orbital motion.\footnote{The direction of the planet's motion does not affect the disk
structure in this model, since a planet's secular perturbations are equivalent to those from the
ring formed by spreading the planet's mass along its orbit (Murray \& Dermott 1999).}
This spiral starts off close to the planet's orbit but propagates out with time.
At its outermost edge the spiral is radially broader than closer to the planet.
This edge generally starts in the pericentre direction, although the densest part
of the spiral is $\sim 180^\circ$ away from that edge closer to the apocentre
direction.
At $0.1t_{sec(3:2)}$ the spiral appears to wrap almost $360^\circ$ around the star.
By $10t_{sec(3:2)}$ at least two windings are evident.
However, the extent to which the inner windings are noticeable in the images is
limited by the resolution of those images.
The spiral structure is more tightly wound both at earlier times and closer to
the planet, and such windings can become too close together to discern.
Where this occurs the disk exhibits a large scale asymmetry which is equivalent
to the pericentre glow approximation (Wyatt et al. 1999) and which is discussed
in greater detail in \S \ref{ss:pgapprox}.
The region covered by the pericentre glow approximation grows with time and by
$100t_{sec(3:2)}$ extends out to $\sim 2.5a_{pl}$ outside of which narrow tightly
wound spiral structure is still apparent.
The evolution is similar for all eccentricities of the planet, although there
are subtle differences, notably the spiral is more open
and the density contrast in and out of the spiral greater for higher
eccentricities.

A similar response is also seen for the inner disk in that a spiral forms which
propagates (inward) away from the planet.
Again the spirals are more tightly wound closer to the planet and at late times the
overlapping spirals cause a large-scale pericentre glow asymmetry.
However, there is an important difference:
the spiral unwinds in the opposite sense to the exterior
spiral (i.e., in the same direction as the planetesimals' motion).
The structure also propagates inward much faster than the outer spiral propagates
outward, and the inner spiral is more open than the outer one.

\begin{figure}
  \centering
    \includegraphics[width=3in]{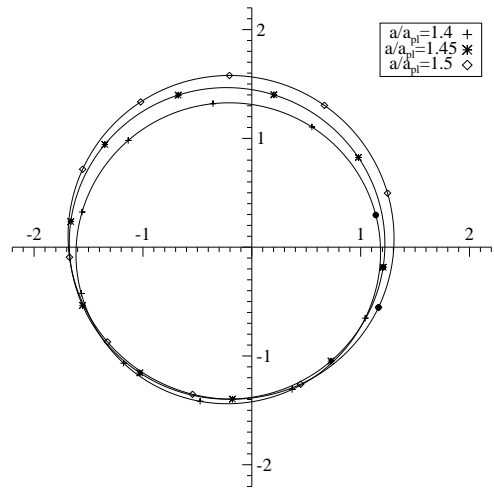}
  \caption{The orbits of planetesimals at semimajor axes of 1.4, 1.45 and
  $1.5a_{pl}$ at a time of $1t_{sec(3:2)}$ after a planet is introduced
  with $e_{pl}=0.1$.
  The x direction denotes the orientation of the planet's pericentre
  and the axes are scaled in units of the planet's semimajor axis.
  The 10 symbols for each orbit are distributed at equal timesteps
  around the orbit with the pericentre denoted by a filled circle.
  Orbital motion is anticlockwise.}
  \label{fig:orbitsfew_epl0.1}
\end{figure}

These distributions were derived solely from the orbital element distributions
shown in Fig.~\ref{fig:ewvsa}.
However, the connection is not immediately apparent, since the orientation of
the spatial intensity maximum at a given distance from the star does not necessarily
correspond to planetesimals which are either at pericentre or apocentre;
note that the spiral winds all the way around the star, whereas
the planetesimals' pericentres are constrained to lie within $90^\circ$
of that of the planet regardless of their distance from the planet.
The origin of the spiral structure is evident from Fig.~\ref{fig:orbitsfew_epl0.1} where
the orbits of planetesimals at $a=1.4, 1.45$ and $1.5a_{pl}$ are plotted at a
time of $1t_{sec(3:2)}$ (see Fig.~\ref{fig:zevolfew} for their eccentricities
and pericentre orientations at this time), assuming that the planet has an
eccentricity of 0.1.
This figure shows how the orbits, despite having different semimajor axes,
eccentricities and pericentre orientations,
are aligned between longitudes of $190-320^\circ$ where the radial width
between the orbits is $<0.05a_{pl}$ compared with the width of $0.25a_{pl}$
at a longitude of $100^\circ$.
The region of aligned orbits in this example corresponds to the parts of the orbits
where the planetesimals are all somewhere between apocentre and pericentre (as the
pericentres are within $25^\circ$ of the x direction, see Fig.~\ref{fig:zevolfew}), and
it is evident from the $1t_{sec(3:2)}$, $e_{pl}=0.1$ panel of Fig.~\ref{fig:pgspimn}
that the alignment of the orbits of such planetesimals is the cause of the spiral
structure.

\begin{figure}
  \centering
    \includegraphics[width=3in]{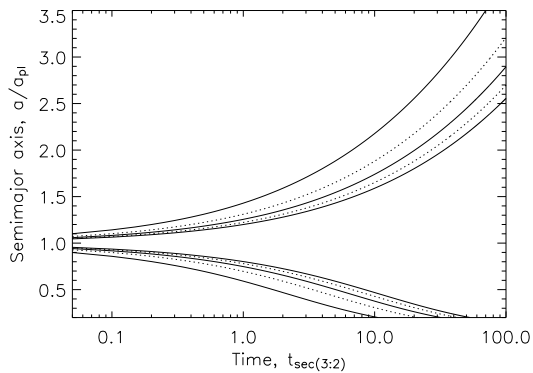}
  \caption{The semimajor axes of planetesimals which have completed
  0.5, 1.5, and 2.5 precession periods (solid lines) and 1.0 and 2.0
  precession periods (dotted lines) at any given time after the
  introduction of an eccentric planet into a disk;
  lower numbers of precession periods correspond to planetesimals further
  from the planet.
  Times are given relative to the secular precession time at the 3:2 resonance,
  $t_{sec(3:2)}$.
  The most prominent spiral in the images is associated with planetesimals
  which have completed half a precession period.
  The spirals from planetesimals at 1.5 and 2.5 precession periods are
  so close that they often appear as an asymmetric ring.}
  \label{fig:aspiralvst}
\end{figure}

The connection between the dynamical and spatial distributions can be further
clarified by considering that the outermost spiral in the outer disk
(or the innermost spiral for the inner disk) is associated with planetesimals
that have completed half a precession period (and so have maximum eccentricity).
The next density maximum is associated with planetesimals that have completed one
and a half precession periods, and so on.
Thus the radial extent of the spiral at different times can be crudely estimated from
Fig.~\ref{fig:aspiralvst} which shows the semimajor axes of planetesimals which have
completed 0.5, 1, 1.5, 2, and 2.5 precession periods at any given time.
This figure also explains many of the features of the spirals in the models, such as why
they are more tightly wound closer to the star and why the inner spiral propagates
faster than the outer spiral.

%%%%%%%%%%%%%%%%%%%%%
\subsection{Pericentre Glow Revisited}
\label{ss:pgapprox}
It is noticeable in Fig.~\ref{fig:pgspimn} that at late times the disk of
planetesimals outside the planet not only has an offset centre of symmetry
(with the forced pericentre side being closer to the star), but that there
is also a density asymmetry in which the forced apocentre side has a higher
peak surface density than the forced pericentre side.
The converse is true for planetesimals inside the planet's orbit
(Fig.~\ref{fig:pgspimnint}), which has its centre of symmetry offset in the
same direction, but the higher density this time is on the forced pericentre
side.
Since the discussion in Fig.~2 of Wyatt et al. (1999) implied that the effect
of a forced eccentricity is only to cause a shift in the centre of symmetry
of the disk, and not in its density, this discrepancy should be cleared up.

The discrepancy is caused by the fact that Fig.~2 of Wyatt et al. (1999)
considered only planetesimals at the same semimajor axis and with the same
forced and proper eccentricities.
If this discussion had included planetesimals at a range of semimajor axes, but again
with the same eccentricities, the result would have been a disk in which the peak surface
densities in the apocentre and pericentre directions are the same, but the ring would
have been broader in the apocentre direction;
this was the assumption used in the modelling of Wyatt et al. (1999).
The change in density around the ring at late times in
Figs.~\ref{fig:pgspimn} and \ref{fig:pgspimnint} is caused by the
fact that the forced eccentricity imposed on the disk is lower for planetesimals that are
further from the planet.
This means that the apocentric contributions from planetesimals at different
semimajor axes exterior to the planet are more bunched up than their equivalent pericentric
contributions, resulting in a disk with a denser peak surface brightness in the forced apocentre
direction, but one in which the width of the ring in the apocentre direction is not as broadened
with respect to the pericentre direction as would otherwise have been the case.
The converse is also true for planetesimals orbiting interior to the planet for which the apocentre
side is even more diffuse and the pericentre side more bunched up than in a disk with a constant
forced eccentricity.

This discussion also raises the issue of apocentre glow, since there is always more
material in the apocentre direction in these models.
The model of Wyatt et al. (1999) found the pericentre side to be brighter because
the increased temperature of material that is closer to the star on the pericentre
side more than compensated for the lesser quantities of dust there.
However, this was the case only because the model was being compared with mid-IR
observations of the HR4796 disk which are very temperature dependent because the
grains are emitting on the Wien side of the black body curve which meant that
disk flux is the product of dust cross-sectional area and a factor
$\propto r^{-2.6}$.
Observations at different wavelengths, such as optical images of scattered light,
or sub-mm images of thermal radiation are less sensitive to dust temperature
(or equivalently distance from the star), since the equivalent multiplicative
factors are $r^{-2}$ for scattered light and $r^{-0.5}$ for sub-mm observations
(because the dust emits on the Rayleigh-Jeans side of the black body curve).
This means that the apocentre side of a disk perturbed by an eccentric companion
could appear brighter in such observations, although a detailed study of this
issue is not appropriate for this paper.

%%%%%%%%%%%%%%%%%%%%%
\subsection{Full Numerical Integration}
\label{ss:full}
\begin{figure}
  \centering
  \begin{tabular}{rl}
    \textbf{(a)} & \includegraphics[width=3in]{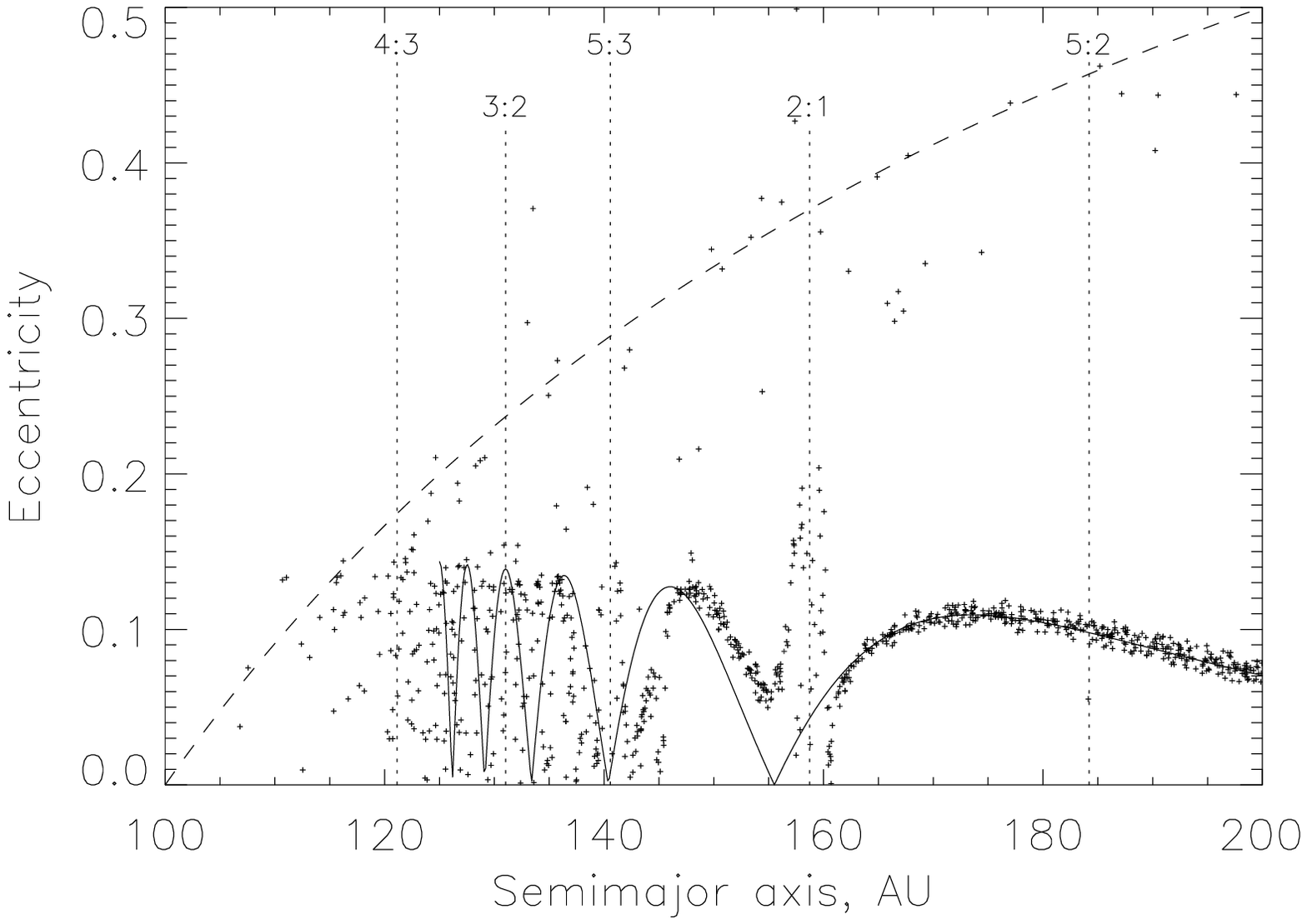} \\
    \textbf{(b)} & \includegraphics[width=3in]{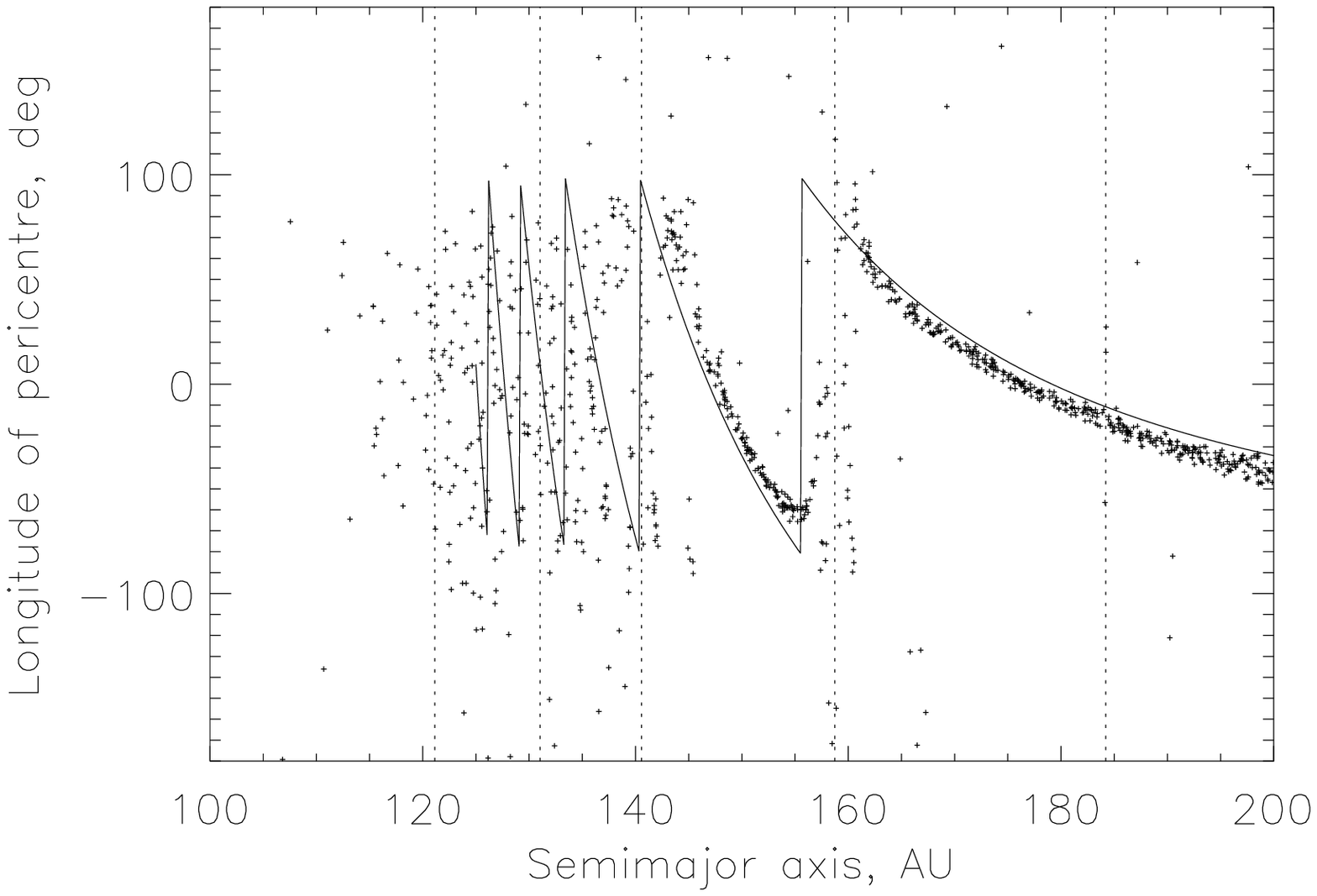}
  \end{tabular}
  \caption{Orbits of planetesimals determined using a full numerical
  integration involving 1000 planetesimals distributed between 110 and 200 AU
  from a $2.5M_\odot$ star:
  \textbf{(a)} eccentricities, and
  \textbf{(b)} pericentre orientations.
  The planetesimals were started on circular orbits and a $80M_\oplus$ planet
  was introduced at 100 AU with an eccentricity of 0.08.
  These plots show the distribution at 15 Myr ($\sim 3.5t_{sec(3:2)}$).
  The solid lines show the distribution predicted from the analytical theory
  and the dotted lines show the locations of the planet's resonances.
  The dashed line delineates planet-crossing orbits.}
  \label{fig:fullew}
\end{figure}

To determine how well the simple first order theory derived above corresponds to
what actually happens, a full numerical integration was performed using a
RADAU fifteenth order integrator (Everhart 1985).
In this integration, the star and planet parameters were set to
$M_\star = 2.5M_\odot$, $M_{pl} = 80M_\oplus$, $e_{pl} = 0.08$, $a_{pl} = 100$ AU,
and 1000 (massless) planetesimals were distributed within the range 110-200 AU with
low eccentricities and inclinations ($e<0.01$, $I<0.573^\circ$).
The orbits of the planetesimals were then integrated due to the gravitational
forces of the star and the planet.
The resulting eccentricity and pericentre distributions after
15 Myr ($\sim 3.5t_{sec(3:2)}$) are shown in Fig.~\ref{fig:fullew}.
Comparison with the analytically determined eccentricities and pericentre orientations
is very good for $a>140$ AU, although the precession rates are slightly faster than
predicted at $a<140$ AU.
Additional perturbations at the locations of the planet's mean motion resonances (e.g., the
2:1 resonance) also lead to deviations from the predicted distribution.
However, these do not affect the overall structure of the planetesimal disk which does
exhibit the spiral structure expected from Fig.~\ref{fig:pgspimn}, since the majority of
this structure is attributable to planetesimals outside 140 AU.
This supports the use of the analytical orbital element distribution to
characterise the spiral structure caused by an eccentric planet, as is the
case in all sections of this paper except this one.
It would be worth considering the effect of these resonant perturbations on the
structure of a disk in more detail in a future work, although this is
beyond the scope of this paper.

\clearpage

%%%%%%%%%%%%%%%%%%%%%
\subsection{Gap Near Planet}
\label{ss:gap}
\begin{figure}
  \centering
  \includegraphics[width=3in]{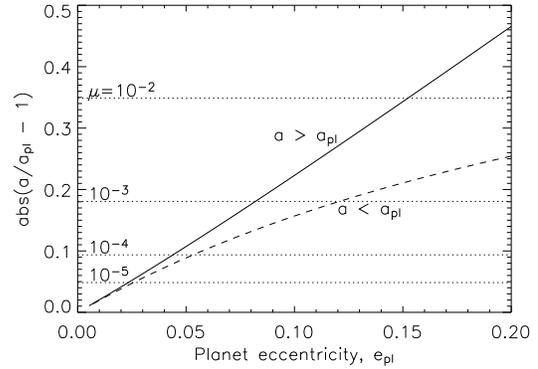}
  \caption{Width of the empty region, $abs(a/a_{pl}-1)$, caused by
  planets of different eccentricities and masses.
  Planetesimals on planet crossing orbits are shown with a solid line for
  external orbits and a dashed line for internal orbits.
  Planetesimals on chaotic orbits due to resonance overlap are shown with dotted
  lines for planet masses defined by $\mu=M_{pl}/M_\star=10^{-2}$, $10^{-3}$,
  $10^{-4}$, and $10^{-5}$.}
  \label{fig:gapvsepl}
\end{figure}

One further consideration for the dynamical structure of the disk not considered
in the previous sections is that planetesimals orbiting near the
planet will be removed from the system on short timescales.
There are two mechanisms which achieve this: 
\textbf{(i)} planetesimals which are on planet crossing orbits,
i.e. those with $a(1-e) < a_{pl}$ and $a(1+e) > a_{pl}$,
will be scattered out after a close encounter with the planet;
\textbf{(ii)} planetesimals in the region where the planet's first
order mean motion resonances are overlapping,
i.e. those with $abs(a/a_{pl}-1) < 1.3 \mu^{2/7}$, have chaotic orbits
and are removed from this region within 1000 encounters (Wisdom 1980).
This results in a gap close to the planet which is devoid of planetesimals.
The sizes of the empty regions caused by mechanisms \textbf{(i)} and \textbf{(ii)} 
are plotted in Fig.~\ref{fig:gapvsepl} using the understanding that at any given
semimajor axis, $e_{max} = 2e_f$.
In practise just one of the mechanisms determines the size of the gap depending
on the eccentricity and mass of the planet.
These planetesimals were not removed from Figs.~\ref{fig:ewvsa}, \ref{fig:pgspimn}
and \ref{fig:pgspimnint} to allow these figures to be completely general with
respect to the planet's mass (and eccentricity).
Removing these planetesimals from Figs.~\ref{fig:pgspimn} and \ref{fig:pgspimnint}
would not change the structure, but would eat away at the edge of the torus closest
to the planet.

%%%%%%%%%%%%%%%%%%%%%
\subsection{Limitations}
\label{ss:limit}
It is worth questioning whether the initial conditions described in
\S \ref{ss:evcold} are realistic.
One concern is the origin of the planet's eccentricity.
Possible mechanisms which act to increase a planet's eccentricity
have been discussed by various authors and include:
scattering of the planet out from closer to the star in an interaction
with other planets which formed there (Thommes et al. 1999; Marzari \&
Weidenschilling 2002);
interaction of the planet with a proto-planetary disk (Goldreich \& Sari 2003;
Papaloizou, Nelson, \& Terquem 2004);
perturbations from stellar encounters (Zamaska \& Tremaine 2004);
and acceleration perturbations caused by asymmetric stellar jets or star-disk
winds (Namouni 2005).
Regardless of the origin of the eccentricity, it is clear from the results
of radial velocity studies that a high proportion
of extrasolar planets do have high eccentricities ($e>0.1$; Fischer et al. 2004),
but this work also shows that only a moderate eccentricity similar
to that of the giant planets in the Solar System ($e \approx 0.05$)
is required to impose significant spiral structure on a disk
(Figs.~\ref{fig:pgspimn} and \ref{fig:pgspimnint}).
Where the model is most limited in this regard is the timescale
over which the planet obtained its mass and eccentricity, and the impact
that had on neighbouring disk material (i.e., whether this material also
received a higher eccentricity in this process), since this model
explicitly refers to the instantaneous introduction of an eccentric planet
into a dynamically cold planetesimal disk.
In this model, instantaneous means fast relative to the timescale
over which secular perturbations act (eq.~\ref{eq:tsec}), thus favouring
mechanisms which introduce an eccentric planet very quickly such as scattering
of a planet from closer in to the star.
Also, further simulations were performed to ascertain the influence of a non-zero
initial planetesimal eccentricity on the resulting spiral structure.
It was found that the same spiral structure is produced as that shown in
Figs.~\ref{fig:pgspimn} and \ref{fig:pgspimnint}, but that the structure is
smoothed out, and the density contrast around the spiral reduced, as the mean
initial planetesimal eccentricity is increased.
The dynamically cold condition can be stated more quantitatively that the
mean initial planetesimal eccentricity must be less than the planet's
eccentricity.

Another limitation of this model is that it is only applicable in
disks that are essentially massless.
The distinction arises because the orbit of a planetesimal will also
be affected by the secular perturbations of neighbouring planetesimals.
If these perturbations are similar in magnitude to those from the planet
they must be taken into account, which I estimate to occur when the
disk is comparable in mass to the planet.
Hahn (2003) described an analytical model which can be used to
determine the secular evolution of such massive disks in which
spiral structure is also seen.
Further, since this model is based on low order secular perturbation theory
(Murray \& Dermott 1999), it is only applicable when eccentricities
are relatively low ($e \ll 1$).

%%%%%%%%%%%%%%%%%%%%%
\section{Application to HD141569}
\label{s:hd}

%%%%%%%%%%%%%%%%%%%%%
\subsection{Observed disk structure}
\label{ss:obsstr}
\begin{figure}
  \centering
  \begin{tabular}{rl}
    \textbf{(a)} & \includegraphics[width=2.0in]{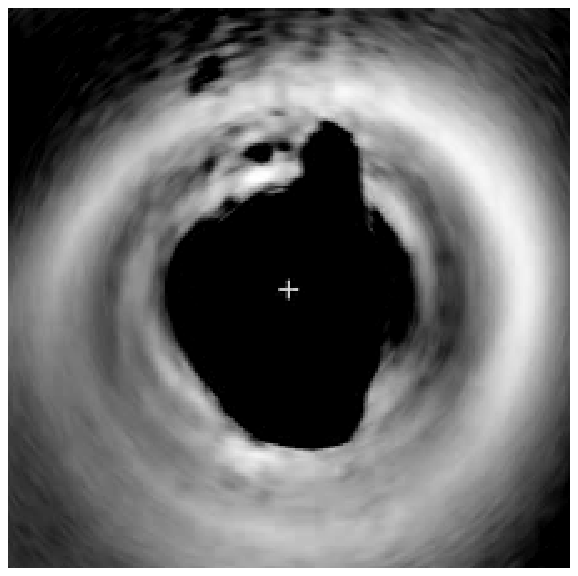} \\[-0.00in]
    \textbf{(b)} & \hspace{-0.8in} \vspace{-0.0in} \includegraphics[width=3.09in]{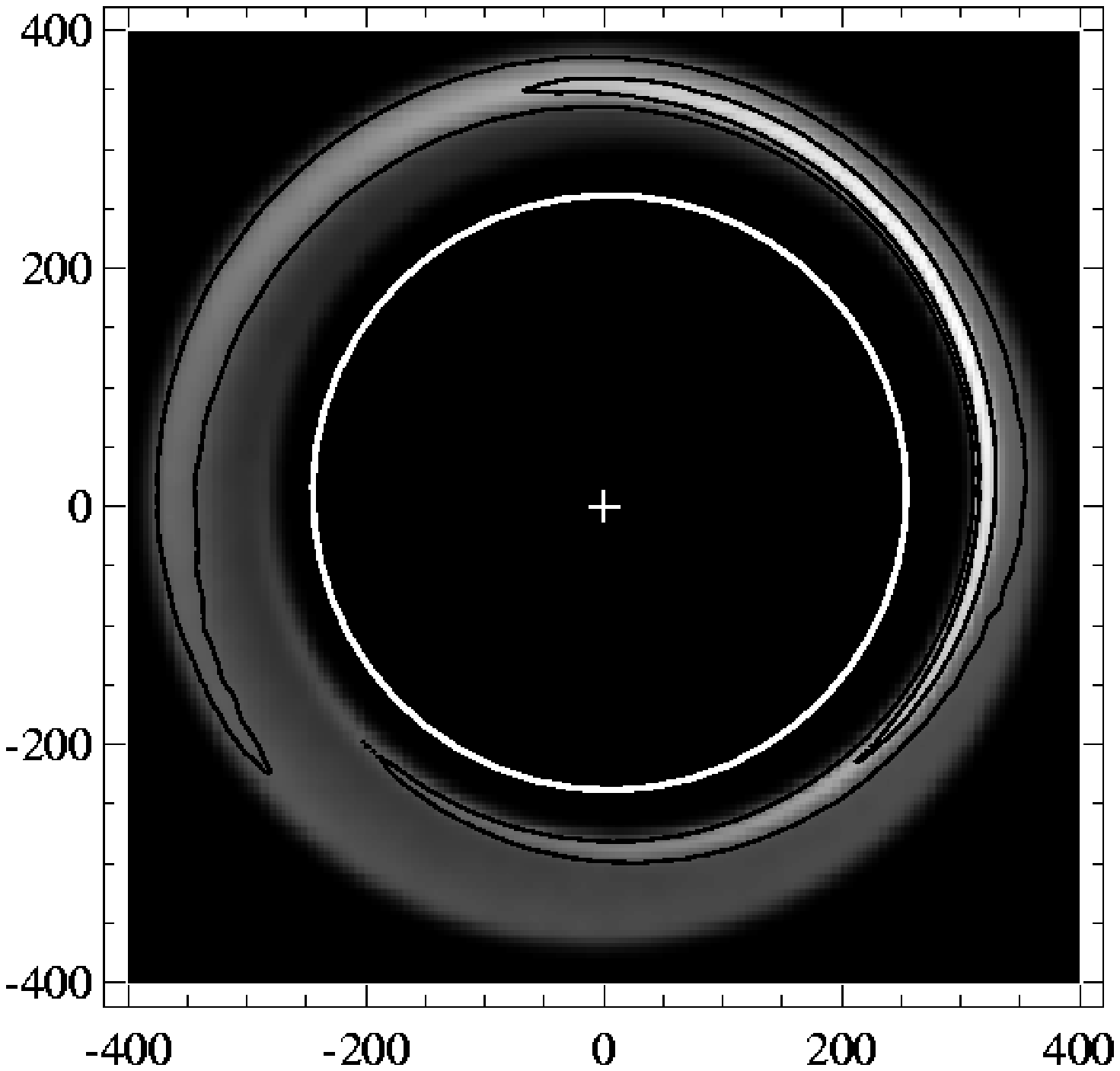} \\[-0.15in]
    \textbf{(c)} & \hspace{-0.8in} \vspace{-0.0in} \includegraphics[width=3.09in]{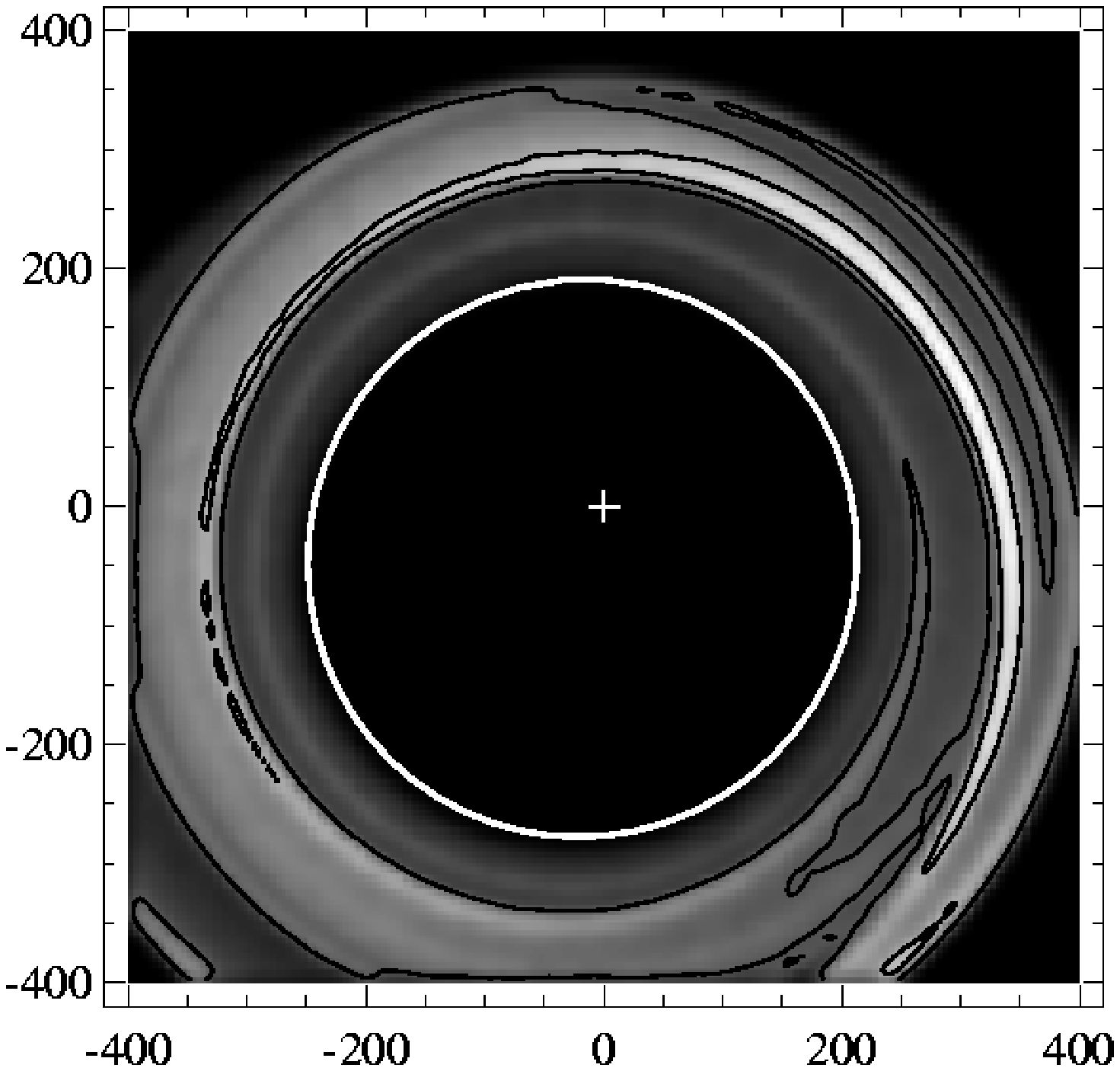}
  \end{tabular}
  \caption{Surface density image of the HD141569 disk:
  \textbf{(a)} observed density distribution reproduced from Fig. 8b of
  Clampin et al. (2003), but cropped so that the length of each side measures
  800 AU;
  \textbf{(b)} model density distribution \textbf{(i)} which fits the spiral at 325 AU
  assuming its structure is as outlined on Fig. 8a of Clampin et al., but
  assumes the star is at the centre of the disk and not where it was
  inferred to be from the images (the plus in \textbf{(a)});
  \textbf{(c)} model density distribution \textbf{(ii)} which fits the spiral
  at 325 AU assuming that the position of the star is as inferred by Clampin
  et al. (2003).
  The two models \textbf{(i)} and \textbf{(ii)} are described in more detail in
  the text.
  The planet's orbit in these models is shown with a white line, the axes are in AU,
  and the position of the star shown with a plus.
  The orientation of all images is such that N is to the left, E is down.}
  \label{fig:hd141569}
\end{figure}

The resolved structure of the HD141569 disk was discussed in \S \ref{s:intro}.
The features which could possibly be explained by the model presented in \S \ref{s:sp}
are those seen in Fig. 8 of Clampin et al. (2003) which shows the image of surface
density of $\mu$m-sized dust in the disk derived from their ACS observations
after deprojection and correction for scattering.
The central 400 AU of this image is reproduced in Fig.~\ref{fig:hd141569}a.
Clampin et al. noted the following features:
tightly wound spiral structures at 200 and 325 AU which unwind in an anticlockwise
direction, both unwinding with a factor of $\sim 1.25$ in distance from
the star between adjacent windings;
gaps at 250 AU and $<175$ AU;
and open spiral arms at $>400$ AU which are not noticeable in the cropped version of
Fig.~\ref{fig:hd141569}a, but which extend from the NW (top left) and SE
(bottom right) portions of the disk in an anticlockwise direction.
Clampin et al. also noted that the star is offset by $\sim 30$ AU from the
centre of the disk toward the W, but quoted an uncertainty of possibly
exceeding $\sim 10$ AU in this offset.

Interpretation of the spiral structures is complicated to some extent
by their observed structure.
One problem is that both the 200 and 325 AU spirals have large azimuthal
density variations which make them appear clumpy.
The 200 AU spiral is so close to the star and the coronagraph that was used to
remove the stellar contribution that its structure may be significantly
affected by uncertainties in the PSF.
It is hard to assess the impact this has on the true structure and in this paper
I assume this to be as outlined on Fig. 8a of Clampin et al., but note that
one should be cautious when attaching too much significance to inferences
based on this structure.
The clumpiness of the 325 AU spiral is particularly noticeable in the E (bottom),
and its structure is further complicated by the presence of the
more distant open spirals.
One issue is that these are likely to affect the apparent radial offset
and density of the 325 AU spiral, particularly in the directions where
the two features meet in the NW and SE.
There is also the question of how the mechanisms which produced the
two spirals interact.
Another problem for all spirals is how to interpret the observed location of
the star.
Naively one might expect the spirals to increase in radius as they unwind.
However, taking the brightest quadrant of the 325 AU spiral as an example, this
appears to peak at 290 AU in the W (top) and 350 AU in the S (right), thus
unwinding in the clockwise direction, even though when traced all the way
around the star the spiral appears to unwind anticlockwise;
the same spiral also peaks at 340 AU in the N (left).
Given these uncertainties I do not rule out that the spiral at 325 AU actually
unwinds clockwise and that its eastern (bottom) portion is affected by both
its clumpiness and the open spirals.

%%%%%%%%%%%%%%%%%%%%%
\subsection{Validity of model}
\label{ss:validity}
Before considering how these features might be explained by the model
of \S \ref{s:sp},
the validity of comparing the structures seen in Fig.~\ref{fig:hd141569}a 
with those in Figs.~\ref{fig:pgspimn} and \ref{fig:pgspimnint}
must be considered.
The first issue to be considered is that the observational evidence points to
the dust grains seen in the scattered light images being small, roughly $\mu$m in size
(Boccaletti et al. 2003; Augereau \& Papaloizou, 2004), similar to inferences
about the minimum grain size from the disk's thermal emission spectrum
(Fisher et al. 2000; Li \& Lunine 2003).
This indicates that the dust is likely to be small enough that it is
in the process of being blown out of the system by radiation pressure.
While the models presented in \S \ref{s:sp} only refer explicitly to the
distribution of planetesimals which are unaffected by radiation pressure,
the short lifetime of $\mu$m-sized dust means that it must originate in the
destruction of such planetesimals.
Since regions where planetesimals are more densely concentrated would have a higher
production rate of small dust grains, which would then be accelerated
out of the system on hyperbolic orbits by radiation pressure,
the morphology of the dust disk seen in scattered light (and in particular
the location of any spiral structure) should be broadly similar to that of
the planetesimals.
This is confirmed by the modelling of the HD141569 disk by Ardila et al. (2005),
who found the surface density distribution of small dust grains in their models
to indeed be very similar to that of the planetesimals, even when radiation
pressure, Poynting-Robertson drag and gas drag are taken into account.
It is, however, contrary to models which explain the wavelength dependence
of the structure of disks such as that around Vega (Su et al. 2005) as a
result of planetesimals and dust grains having different distributions
(Wyatt et al. 2005).

It is also worth considering if the disk is massive enough to need to consider
the secular perturbations from the disk itself (Hahn 2003).
The dust disk has a mass of $\sim 0.4M_\oplus$ (Sheret et al. 2004), indicating
that its self-gravity can be ignored (assuming an Earth mass planet or larger).
However, the unseen planetesimal disk which feeds the dust disk could be more
massive than this (e.g., Wyatt \& Dent 2002) in which case the effect of
self-gravity might need to be considered.

In other words there are still uncertainties about exactly how
well the structures seen in the models would describe the observed structure
of the HD141569 disk.
However, it is not the aim of this paper to provide a complete description
of the HD141569 disk.
The disk's structure is not likely to be solely attributable to the
secular perturbations of planets on eccentric orbits, rather more than
one process is probably responsible, including the secular perturbations and
tidal forces from the binary companion (Augereau \& Papaloizou 2004; Quillen
et al. 2005) and those of embedded planets (this work), and possibly also
from the resonant perturbations of those objects (e.g., Kuchner \& Holman
2003; Wyatt 2003).
Rather in the spirit of Augereau \& Papaloizou (2004) the aim of this paper
is to show which of its features might (or might not) be attributable to
the secular perturbations of a low eccentricity planet.

%%%%%%%%%%%%%%%%%%%%%
\subsection{Interpretation of disk structure}
\label{ss:interp}
It is important to start by ruling out the structures which cannot be
explained by the secular perturbations of an eccentric planet.
The spiral structure imposed on a disk outside a planet is tightly wound so
that the outermost spiral is limited in terms of how distant it can be to
the one interior to it (e.g., Fig.~\ref{fig:aspiralvst}).
Thus the open spiral arms at $>400$ AU cannot be explained by a low
eccentricity ($e \leq 0.2$) internal planet.
However, since the spirals interior to a planet can be more open
this suggests that the open spirals are more likely to be associated with an
external perturber such as the binary companion as suggested by Quillen
et al. (2005) and Ardila et al. (2005).
Thus the disk's structure at $>400$ AU is not considered further in this
paper, except for the possibility that it may affect the interpretation of
the spiral at 325 AU.

Another possibility that can be ruled out straight away is that there is a
planet orbiting in the gap at $<175$ AU and that this is responsible for a
spiral which unwinds all the way from 200 to 325 AU, but that there is some
additional mechanism removing material from the region at 250 AU.
This is ruled out because the spirals at 200 and 325 AU unwind at a
similar rate whereas in this model the windings that are closer to the planet
are much closer together than those further away (e.g.,
Fig.~\ref{fig:aspiralvst}).

Here the constraints on planets in the disk are discussed based on two
assumptions about the true disk structure.

\textbf{(i)}
The first assumption is that the 325 AU spiral is as traced in Fig. 8a of
Clampin et al. (2003) and that the location of the star is at the centre of
the disk, rather than offset by 30 AU as inferred (with low significance)
from the images.
Possible planet configurations are constrained most strongly by the tightness
of the windings of the spiral, which overlaps itself with a separation of 80 AU
implying a factor of $\sim 1.25$ between adjacent windings.
This is only compatible with this being the outermost winding
of a planet's outer spiral if that planet was introduced $0.1-0.3t_{sec(3:2)}$
ago with an eccentricity of $<0.1$ (Figs.~\ref{fig:pgspimn} and
\ref{fig:aspiralvst}).
Longer timescales and higher eccentricities result in windings that are too
open, and shorter timescales result in a spiral which does not extend
completely around the star.
If this is the case, the planet would be orbiting close to the inner edge of
spiral at $250-300$ AU, with planetesimals orbiting in a clockwise direction.
Clockwise rotations for the disk and binary companion were also
inferred by the modelling of Quillen et al. (2005) and Ardila et al. (2005)
from the shape of the spiral arms at $>400$ AU;
Augereau \& Papaloizou (2004) did not discuss the sense of these rotations in
their model.

An example of a suitable configuration is shown in Fig.~\ref{fig:hd141569}b.
This figure assumes a planet orbit with $a=250$ AU, $e=0.05$,
and a pericentre orientation in the ENE (just left of down), and planetesimals
on clockwise orbits.
The surface density distribution has also been smoothed by a Gaussian with a
FWHM of 16 AU.
\footnote{While the PSF of the ACS observations had a FWHM of 5 AU, the image in
Fig.~\ref{fig:hd141569}a has also been deprojected and smoothed
azimuthally along $\sim 18$ AU arcs leading to a smoothing function
which varies across the image.
However, the apparently higher resolution of the models in Fig.~\ref{fig:hd141569}
is more likely to be caused by the difference in the distributions of the
planetesimals in the disk (shown in the figures) and those of the dust grains
seen in the observation (see \S \ref{ss:validity}).}
The planet was introduced $0.3t_{sec(3:2)}$ ago, and
material has been excluded that has semimajor axes of $<300$ AU or $>375$ AU;
in other words the gap is slightly larger than would be expected from
\S \ref{ss:gap}, and material further from the star has been assumed to
have been incorporated into the open spiral structure at $>400$ AU
by some other mechanism.
The mass of such a planet can be estimated from the following equation:
\begin{equation}
  M_{pl}/M_{Jup} = 680\frac{N_{sec(3:2)}}{t_{age}}
    \sqrt{ \left( \frac{M_\star}{M_\odot} \right)
           \left( \frac{a_{pl}}{a_\oplus} \right)^3 }, \label{eq:mpl}
\end{equation}
where $N_{sec(3:2)}$ is the number of secular precession periods (eq.~\ref{eq:tsec32})
completed since the planet was introduced $t_{age}$ years ago and $M_{Jup}$ is the mass
of Jupiter.
Assuming the planet was introduced when the star (of mass $2.5M_\odot$) was born
5 Myr ago, this gives a mass of $0.2-0.3M_{jup}$ for the planet in this scenario,
similar to that of Saturn;
the gap caused by resonance overlap by such a planet would be similar in size
to that created by scattering of planetesimals on planet-crossing orbits
(Fig.~\ref{fig:gapvsepl}).
Higher mass planets would be required if the planet was introduced more
recently (and would clear wider gaps);
e.g., the gap used in Fig.~\ref{fig:hd141569}b, if caused by resonance overlap, implies a
$\sim 3M_{Jup}$ planet, which would have to have been introduced $\sim 0.5$ Myr
ago to cause the same structure.

It is also possible that the spirals are those associated with planetesimals
that have completed $\geq 1.5$ precession periods, in which case some mechanism
must be invoked to disperse the outermost spiral as well as to remove the
material close to the planet which would otherwise cause a dense region of
pericentre glow.
Neither of these conditions is unreasonable as the outer edge of the disk would
be affected by the same mechanism which causes the open spirals, and mechanisms
exist which clear material close to the orbit of a planet (\S \ref{ss:gap}).
This scenario would require a planet with a mass much higher than that of
Saturn due to the higher value of $N_{sec(3:2)}$ (Fig.~\ref{fig:pgspimn};
eq.~\ref{eq:mpl}).
However, I consider this possibility unlikely because the planet would also
have to be orbiting at $\ll 250$ AU, which would mean that the gap it causes
along its orbit would overlap the dense ring observed at 200 AU. 

An appealing prediction of this interpretation would be if the same planet
is causing the spiral observed at 200 AU.
However, this cannot be the case if that spiral also unwinds anticlockwise,
since the model would predict it to unwind in the clockwise direction.
Should subsequent observations find the spiral at 200 AU to unwind clockwise,
this would lend support to this interpretation of the disk structure.
However, for now the spiral at 200 AU is assumed to unwind anticlockwise.
Since Clampin et al. found this spiral to be wound with a similar
tightness to that at 325 AU, the model shown in Fig.~\ref{fig:hd141569}b is
equally applicable when appropriately scaled to the spiral at 200 AU
(ignoring the effect of any perturbations from a possible planet at 250 AU).
That is, the structure at 200 AU could be caused by a planet orbiting
at $\sim 150$ AU with $e=0.05$ and a mass about half that of the inferences
about the planet at 250 AU because of its proximity to the star
(eq.~\ref{eq:mpl}).
However, it is not possible to rule out that the planet is actually
much closer to the star as it was for the planet causing the spiral at
325 AU, because the inner structure of the disk is poorly known at
present.

\textbf{(ii)}
The second assumption is that the offset from the star was determined
accurately from the observations, but that the spiral is not well
characterised in the E due to the clumpiness of the disk and confusion
with the open spiral structure at $>400$ AU.
Thus the aim is for the model to reproduce offsets of 340, 290 and
350 AU in the N, W and S directions;
the constraint in the E is that the spiral should peak between 350
and 430 AU, depending on how it traverses the N-W quadrant
(or vice versa).
There is a class of spirals in this model that do fit the specification
that the radial offset does not increase monotonically as the spiral
unwinds.
This occurs at late times ($>1t_{sec(3:2)}$) and is most noticeable for
high eccentricity ($e>0.15$) planets (see Fig.~\ref{fig:pgspimn}).
It is caused by the spiral having its centre of symmetry offset from
the star in the direction of the planet's apocentre by an amount that
is proportional to the forced eccentricity imposed on the ring
(e.g., Wyatt et al. 1999).
This does not occur at earlier times, because the material seen in
outermost part of the spiral in the pericentre direction has not
precessed far enough around the forced eccentricity
(Fig.~\ref{fig:zevolfew}).

An example of a suitable configuration is shown in Fig.~\ref{fig:hd141569}c.
This figure assumes a planet orbit with $a=235$ AU, $e=0.2$,
and a pericentre orientation in the WSW (just right of up).
This time planetesimals in the disk are orbiting anticlockwise.
Again, additional smoothing by a Gaussian with a FWHM of 16 AU was
applied to the resulting surface density distribution.
The planet was introduced $3t_{sec(3:2)}$ ago, and
material has been excluded that has semimajor axes of $<294$ AU or $>423$ AU;
this time the gap is slightly larger than would be expected from
\S \ref{ss:gap}.
The mass of such a planet must be at least $2M_{Jup}$ (eq.~\ref{eq:mpl};
more if the planet was introduced more recently than 5 Myr ago).
This model fits the constraints at all azimuthal angles, even qualitatively
reproducing the structure in the N to E.

The biggest drawback of this interpretation is that the
planet is much closer to the star putting its orbit almost coincident
(though just outside) the material seen in the W and E of the ring at 200 AU.
One possibility is that this is not just a coincidence:
this ring could be comprised of material that is either trapped in
1:1 resonance with the planet;
or this ring is material that has migrated out from closer to the star
due to the combined action of gas drag and radiation pressure to
the edge of the gas disk (Takeuchi \& Artymowicz 2001), a disk
which could have been truncated by the planet.
It should also be pointed out that it is worth considering the effect
of the planet's resonances in models with such high mass planets
(e.g., \S \ref{ss:full}).
This could in fact be beneficial to the model in its application
to the HD141569 disk, since it could explain some of the disk's
clumpy structure (e.g., Wyatt 2003).
Also, anticlockwise planetesimal orbits are contrary to the sense of rotation
inferred for the disk and binary companion by Quillen et al. (2005)
and Ardila et al. (2005).

%%%%%%%%%%%%%%%%%%%%%
\section{Conclusions}
\label{s:conc}
This paper shows how the secular perturbations of a planet on an eccentric
orbit imposes spiral structure on a circumstellar disk after it is introduced
into the system.
A simple model for the spiral structure is presented which is based on the effect of
the planet on the orbits of nearby material.
Two spirals are produced, one exterior and one interior to the planet.
The two spirals unwind in opposite directions.
Disk structure is shown to be dependent only on the time since the planet was
introduced relative to a characteristic secular precession timescale
(which is dependent on the masses of the planet and star and the planet's
semimajor axis), and to a lesser extent on the planet's eccentricity.
The spirals are transient structures that propagate away from the planet. 
At late times (typically after several Myr) the pericentre glow approximation
is recovered in which the disk forms a torus with an offset centre of symmetry
being closer to the star on the side of the planet's pericentre.
This torus is shown to be non-uniform with a higher surface density 
in the apocentric direction.
This could lead to cases of apocentre glow for observations at wavelengths
for which the higher density in the apocentre direction is not counteracted
by the greater distance from the star relative to the pericentre side.

Application of the model to the HD141569 disk shows that some of its structure
may be explained by the presence of planets on eccentric orbits that are 
embedded in its disk.
Two models are presented that explain the spiral structure at 325 AU, one of
which requires a $\sim 0.3 M_{Jup}$ planet orbiting at 250 AU with an 
eccentricity of 0.05, the other requiring a $\sim 3M_{Jup}$ planet orbiting at
235 AU with an eccentricity of 0.2.
The distinction between the models is the significance which is attached to
the location of the star within the disk as derived by Clampin et al. (2003).
The models include large uncertainties, particularly on the mass of the
planet which could be much more massive if the planet formed more recently than
5 Myr ago.
The models also show that the spiral at 200 AU could be explained by a planet
as small as Saturn orbiting at 150 AU.
However, for now this interpretation should not be given much emphasis,
since the disk's structure at $\leq 200$ AU is poorly known at present.

This paper does not purport to provide a complete description of the
HD141569 disk, rather to illustrate the types of structure caused by planets
on moderately eccentric orbits and the way in which observations can be
used to set constraints on planets in disks with detected spiral structure.
Such a model for HD141569 would require a proper treatment of the relation of the
dust distribution to that of the parent planetesimals based on a knowledge of
the disk's solid and gaseous component at hundreds of AU.
It would also have to consider that the disk is likely to have been shaped
by more than one physical process.
In this respect, more accurate knowledge of the disk's structure (in particular
the location of the star within the disk) and of the other perturbations to
which the disk is subjected (e.g., by determining the orbit of the known binary
companion) would allow firmer constraints to be set on any planetary perturbers.
Determining the sense of the disk's orbital motion is also important,
since the two models presented here for the spiral at 325 AU predict motions
in different directions.

However, what these results do show is that the structure of the HD141569 disk
may indicate that there is one, possibly even two, giant planets orbiting
at hundreds of AU from HD141569 on moderately eccentric orbits, and that
constraints can be set on those planets using the disk's observed structure.
If this interpretation turns out to be true, and such planets do exist,
this would set important constraints on the mechanism through which massive
planets form, or come to be, at such large distances, particularly given
the relatively young age of this system of $\sim 5$ Myr and the knowledge of
the structure of the disk near the planet.

%\begin{acknowledgements}
%We gratefully acknowledge...
%\end{acknowledgements}

%%%%%%%%%%%%%%%%%%%%%

\end{document}